\documentclass[11pt]{article}
\usepackage{amssymb,amsmath,amsfonts}
\usepackage{graphicx}
\usepackage{graphics}
\usepackage{eepic,epsfig}

\begin{document}

\title{Vacuum polarization by a composite topological defect}
\author{E. R. Bezerra de Mello$^{1}$\thanks{%
E-mail: emello@fisica.ufpb.br}\, and A. A. Saharian$^{1,2}$\thanks{%
E-mail: saharyan@server.physdep.r.am} \\
\\
\textit{$^1$Departamento de F\'{\i}sica-CCEN, Universidade Federal da Para%
\'{\i}ba}\\
\textit{58.059-970, Caixa Postal 5.008, Jo\~{a}o Pessoa, PB, Brazil}\vspace{%
0.3cm}\\
\textit{$^2$Department of Physics, Yerevan State University,}\\
\textit{375025 Yerevan, Armenia}}
\maketitle

\begin{abstract}
In this paper we analyze one-loop quantum effects of a scalar field induced
by a composite topological defect consisting a cosmic string on a $p$%
-dimensional brane and a $(m+1)$-dimensional global monopole in the
transverse extra dimensions. The corresponding Green function is presented
as a sum of two terms. The first one corresponds to the bulk where the
cosmic string is absent and the second one is induced by the presence of the
string. For the points away from the cores of the topological defects the
latter is finite in the coincidence limit and is used for the evaluation of
the vacuum expectation values of the field square and energy-momentum tensor.
\end{abstract}

\bigskip


\bigskip

\section{Introduction}

In recent years the braneworld model has received renewed interest. By this
scenario our world is represented by a sub-manifold, a three-brane, embedded
in a higher dimensional spacetime (for a review see \cite{Ruba01,Maar04}).
Braneworlds naturally appear in the string/M theory context and provide a
novel setting for discussing phenomenological and cosmological issues
related to extra dimensions. The models introduced by Randall and Sundrum
are particularly attractive \cite{RS,RS1}. The corresponding spacetime
contains two (RSI), respectively one (RSII), Ricci-flat brane(s) embedded on
a five-dimensional Anti-de Sitter (AdS) bulk. It is assumed that all matter
fields are confined on the branes and only the gravity propagates in the
five dimensional bulk. More recently, alternatives to confining particles on
the brane have been investigated \ and scenarios with additional bulk fields
have been considered.

Although topological defects have been first analyzed in four-dimensional
spacetime \cite{VS}, they have been considered in the context of braneworld.
In this scenario the defects live in a $d-$dimensional submanifold embedded
in a $(4+d)-$dimensional Universe. The domain wall case, with a single extra
dimension, has been considered in \cite{Rubakov}. More recently the cosmic
string case, with two additional extra dimensions, has been analyzed \cite%
{Cohen,Ruth}. For the case with three extra dimensions, the 't
Hooft-Polyakov magnetic monopole has been numerically analyzed in \cite%
{Roessl,Cho}. In Refs. \cite{Ola}-\cite{Cho2} the corresponding analysis for
the global monopole is presented. In particular, in \cite{Ola} the authors
have obtained the solution to the Einstein equations considering a general $p
$-dimensional Minkowski brane worldsheet and a $(m+1)$-dimensional global
monopole in the transverse extra dimensions with the core on the brane. The
corresponding line element has the form
\begin{equation}
ds^{2}=\eta _{\mu \nu }dx^{\mu }dx^{\nu }-dr^{2}-\alpha ^{2}r^{2}d\Omega
_{m}^{2},  \label{g}
\end{equation}%
where $\eta _{\mu \nu }$ is the Minkowskian metric on the $p$-brane, $\alpha
^{2}=1-\kappa ^{2}\eta _{0}^{2}/(m-1)$ with $\eta _{0}$ being the energy
scale where the gauge symmetry of the global system is spontaneously broken,
$d\Omega _{m}^{2}$ is the line element on a $m$-sphere with unit radius.

In braneworld models the investigation of quantum effects is of considerable
phenomenological interest, both in particle physics and in cosmology. The
analysis of quantum effects produced by a massless scalar field propagating
in the bulk described by $m=2$ version of line-element (\ref{g}) has been
developed in \cite{Mello0} (for the investigation of the quantum vacuum
effects in higher dimensional braneworld models with compact internal spaces
see \cite{Flac03}-\cite{Saha06b}). Continuing in this direction, our
interest here is to investigate the quantum effects induced also by the
conical structure of the brane. We will consider the cosmic string on the
brane and the global monopole in transverse dimensions as idealized defects.
The paper is organized as follows. In the next section we evaluate the heat
kernel for a massless scalar field. In Section \ref{sec:integ} we consider a
special case where the angle deficit in the cosmic string subspace is an
integer fraction of $2\pi $ and evaluate the corresponding Euclidean Green
function and the vacuum expectation values of the field square. The
corresponding quantities for the general case of the angle deficit are
evaluated in Section \ref{sec:gencase}. We also consider the vacuum
expectation value of the energy-momentum tensor.

\section{Heat kernel}

\label{sec:Heat}

We consider a $(D+1)$-dimensional background spacetime having the structure
of direct product of the cosmic string and global monopole subspaces with
the line element
\begin{equation}
ds^{2}=dt^{2}-d\rho ^{2}-\rho ^{2}d\phi
^{2}-\sum_{i=1}^{d}dz_{i}^{2}-dr^{2}-\alpha ^{2}r^{2}d\Omega _{m}^{2}\ ,
\label{ds2comp}
\end{equation}%
where $\phi \in \lbrack 0,2\pi /b]$, $z_{i}\in (-\infty ,\infty )$, and $%
D=d+m+3$. In the standard case with $d=1$, the parameter $b$ is associated
with the planar angle deficit and is related to the linear mass density of
the string. In the special case $b=1$, line element (\ref{ds2comp}) reduces
to the interval (\ref{g}) with $p=d+3$. In this section we evaluate the heat
kernel associated with a massless scalar field in the spacetime defined by (%
\ref{ds2comp}) admitting an arbitrary curvature coupling parameter $\xi $.
The corresponding Euclidean Green function obeys the second order
differential equation
\begin{equation}
\left( \nabla _{l}\nabla ^{l}+\xi R\right) G(x,x^{\prime })=-\delta
^{D+1}(x,x^{\prime })\ ,  \label{B}
\end{equation}%
with $\nabla _{l}$ being the covariant derivative operator and $R$ is the
scalar curvature for the background spacetime. The most important special
cases are minimally and conformally coupled scalar fields with $\xi =0$ and $%
\xi =\xi _{D}\equiv (D-1)/4D$ respectively. For the geometry described by (%
\ref{ds2comp}) one has $R=m(m-1)(1/\alpha ^{2}-1)/r^{2}$ for the points away
from the string core. In order to evaluate the Green function we adopt the
Schwinger-DeWitt formalism as shown below:
\begin{equation}
G(x,x^{\prime })=\int_{0}^{\infty }dsK(x,x^{\prime };s)\ ,  \label{Heat}
\end{equation}%
where the heat kernel, $K(x,x^{\prime };s)$, can be expressed in terms of
eigenfunctions of the operator $\nabla _{l}\nabla ^{l}+\xi R$ as follows:
\begin{equation}
K(x,x^{\prime };s)=\sum_{\sigma }\varphi _{\sigma }(x)\varphi _{\sigma
}^{\ast }(x^{\prime })\exp (-s\sigma ^{2})\ ,  \label{Heat-1}
\end{equation}%
$\sigma ^{2}$ being the corresponding positively defined eigenvalue. Writing
\begin{equation}
\left( \nabla _{l}\nabla ^{l}+\xi R\right) \varphi _{\sigma }(x)=-\sigma
^{2}\varphi _{\sigma }(x)\ ,  \label{eigmod}
\end{equation}%
we obtain the complete set of normalized solutions of the above equation
\begin{equation}
\varphi _{\sigma }(x)=\left( \frac{q\lambda \alpha ^{-m}b}{N(m_{\eta })}%
\right) ^{\frac{1}{2}}\frac{e^{-i\omega \tau +inb\phi +i\mathbf{k}\cdot
\mathbf{z}}}{(2\pi )^{n/2+1}r^{(m-1)/2}}J_{|n|b}(q\rho )J_{\nu _{l}}(\lambda
r)Y(m_{\eta };\vartheta ,\Phi )\ ,  \label{Phisig}
\end{equation}%
with $\tau $ being the Euclidean time coordinate, $\mathbf{z}=(z_{1},\ldots
,z_{d})$, $k=|\mathbf{k}|$, $n=0,\pm 1,\ldots $, and
\begin{equation}
\sigma ^{2}=\omega ^{2}+k^{2}+q^{2}+\lambda ^{2}\ .  \label{sig2}
\end{equation}%
The expression for $N(m_{\eta })$ is given in \cite{Erdelyi} and will not be
necessary in the following discussion. In (\ref{Phisig}), $J_{\nu }(x)$ is
the Bessel function,
\begin{equation}
\nu _{l}=\frac{1}{\alpha }\left[ \left( l+\frac{m-1}{2}\right)
^{2}+(1-\alpha ^{2})m(m-1)\left( \xi -\xi _{m}\right) \right] ^{\frac{1}{2}},
\label{nu}
\end{equation}%
the function $Y(m_{\eta };\vartheta ,\Phi )$ is the hyperspherical harmonic
of degree $l$ \cite{Erdelyi} with $m_{\eta }=(m_{0}\equiv l,m_{1},\ldots
,m_{m-1})$, and $m_{1},m_{2},\ldots ,m_{m-1}$ are integers such that
\begin{equation}
0\leq m_{m-2}\leq m_{m-3}\leq \cdots \leq m_{1}\leq l,\quad -m_{m-2}\leq
m_{m-1}\leq m_{m-2}.  \label{mnumbvalues}
\end{equation}%
So according to (\ref{Heat-1}) the heat kernel is given by the expression
\begin{equation*}
K(x,x^{\prime };s)=\int d\omega \int d^{d}k\int dq\int d\lambda
\sum_{n}\sum_{m_{\eta }}\varphi _{\sigma }(x)\varphi _{\sigma }^{\ast
}(x^{\prime })e^{-s\sigma ^{2}}\ .
\end{equation*}%
By using the formula from \cite{Pru} for the integrals involving
the Bessel functions and the addition theorem for the spherical
harmonics \cite{Erdelyi}
\begin{equation}
\sum_{m_{\eta }}\frac{Y(m_{\eta };\vartheta ,\Phi )}{N(m_{\eta })}Y^{\ast
}(m_{\eta };\vartheta ^{\prime },\Phi ^{\prime })=\frac{2l+m-1}{(m-1)S_{m}}%
C_{l}^{(m-1)/2}(\cos \theta ),  \label{adtheorem}
\end{equation}%
we obtain the following formula
\begin{eqnarray}
K(x,x^{\prime };s) &=&\frac{b\left( rr^{\prime }\right) ^{(1-m)/2}}{(4\pi
)^{p/2}\alpha ^{m}}\frac{e^{-V/4s}}{s^{p/2+1}}\sideset{}{'}{\sum}%
_{n=0}^{\infty }I_{nb}\left( \frac{\rho \rho ^{\prime }}{2s}\right) \cos
(nb\Delta \phi )  \notag \\
&&\times \sum_{l=0}^{\infty }\frac{2l+m-1}{(m-1)S_{m}}C_{l}^{(m-1)/2}(\cos
\theta )I_{\nu _{l}}\left( \frac{rr^{\prime }}{2s}\right) \ ,  \label{Kxx2}
\end{eqnarray}%
where $I_{\nu }(x)$ is the modified Bessel function, $C_{p}^{q}(x)$ is the
Gegenbauer polynomial of degree $p$ and order $q$,
\begin{equation*}
V=\Delta \tau ^{2}+\Delta \mathbf{z}^{2}+\rho ^{2}+\rho ^{\prime
}{}^{2}+r^{2}+r^{\prime }{}^{2}\ ,
\end{equation*}%
and the prime means that the summand with $n=0$ should be taken with the
weight $1/2$. In formula (\ref{Kxx2}), $\theta $ is the angle between the
directions $(\vartheta ,\Phi )$ and $(\vartheta ^{\prime },\Phi ^{\prime })$%
, \ and $S_{m}=2\pi ^{(m+1)/2}/\Gamma ((m+1)/2)$ is the volume of the $m$%
-dimensional sphere. Note that $m=1$ corresponds to a cosmic string in
transverse dimensions. In this case $\nu _{l}=l/\alpha $ and the
corresponding heat kernel is obtained from general formula (\ref{Kxx2}) by
taking into account the relation (see, for instance, \cite{Ab}) $\lim_{\beta
\rightarrow 0}lC_{l}^{\beta }(\cos \theta )/\beta =(2-\delta _{l0})\cos
l\theta $ $\ $\ for the Gegenbauer polynomial.

\section{ Special case}

\label{sec:integ}

Before to construct the Green function in the general case by using (\ref%
{Heat}), we will consider a special case when the parameter $b$ is an
integer number. In order to provide the Green function let us go back,
before to make the integration over $q$. Using the formula \cite{Pru}
\begin{equation}
\sum_{n=-\infty }^{\infty }J_{|n|b}(q\rho )J_{|n|b}(q\rho ^{\prime
})e^{inb\Delta \phi }=\frac{1}{b}\sum_{j=0}^{b-1}J_{0}(qv_{j})\ ,
\label{Sumn}
\end{equation}%
with
\begin{equation}
v_{j}^{2}=\rho ^{2}+\rho ^{\prime }{}^{2}-2\rho \rho ^{\prime }\cos \left(
\Delta \phi -2\pi j/b\right) \ ,  \label{vj}
\end{equation}%
we find
\begin{equation}
K(x,x^{\prime };s)=\frac{\left( rr^{\prime }\right) ^{(1-m)/2}}{2(4\pi
)^{p/2}\alpha ^{m}}\sum_{j=0}^{b-1}\frac{e^{-V_{j}/4s}}{s^{p/2+1}}%
\sum_{l=0}^{\infty }\frac{2l+m-1}{(m-1)S_{m}}C_{l}^{(m-1)/2}(\cos \theta
)I_{\nu _{l}}\left( \frac{rr^{\prime }}{2s}\right) ,  \label{Kxspecial}
\end{equation}%
where
\begin{equation}
V_{j}=\Delta \tau ^{2}+\Delta \mathbf{z}^{2}+r^{2}+r^{\prime
}{}^{2}+v_{j}^{2}\ .  \label{Vj}
\end{equation}

Now we are in position to obtain the Euclidean Green function by
substituting (\ref{Kxspecial}) into (\ref{Heat}). After the evaluation of
the integral by using the formula from \cite{Pru}, our final result is:
\begin{eqnarray}
G(x,x^{\prime }) &=&-\frac{(-i)^{d}(2\pi )^{-d/2-2}}{\alpha ^{m}(rr^{\prime
})^{(D-1)/2}}\sum_{j=0}^{b-1}\frac{1}{(\sinh u_{j})^{d/2+1}}\times   \notag
\\
&&\sum_{l=0}^{\infty }\frac{2l+m-1}{(m-1)S_{m}}C_{l}^{(m-1)/2}(\cos \theta
)Q_{\nu _{l}-1/2}^{d/2+1}(\cosh u_{j})\ ,  \label{Gxxinteg1}
\end{eqnarray}%
where
\begin{equation}
\cosh u_{j}=\frac{V_{j}}{2rr^{\prime }},\   \label{u}
\end{equation}%
and $Q_{\nu }^{\lambda }(x)$ is the associated Legendre function. By making
use of the relation between the Legendre function and the hypergeometric
function \cite{Ab}, formula (\ref{Gxxinteg1}) can also be written in the form%
\begin{eqnarray}
G(x,x^{\prime }) &=&\frac{\alpha ^{-m}}{(2\pi )^{\frac{p}{2}}(rr^{\prime })^{%
\frac{D-1}{2}}}\sum_{l=0}^{\infty }\frac{2l+m-1}{(m-1)S_{m}}\frac{2^{-\nu
_{l}-1}\Gamma (\mu _{l})}{\Gamma (\nu _{l}+1)}C_{l}^{(m-1)/2}(\cos \theta )
\notag \\
&&\times \sum_{j=0}^{b-1}(\cosh u_{j})^{-\mu _{l}}F\left( \frac{\mu _{l}+1}{2%
},\frac{\mu _{l}}{2};\nu _{l}+1;\frac{1}{\cosh ^{2}u_{j}}\right) ,
\label{Gxxinteg2}
\end{eqnarray}%
where we have introduced the notation%
\begin{equation}
\mu _{l}=\nu _{l}+p/2.  \label{mul}
\end{equation}%
From (\ref{Gxxinteg2}) we can observe that the $j=0$ component clearly
presents a divergence at the coincidence limit. However, for the other
components $\cosh u_{j}$ will be always greater than unity and consequently
the Legendre function assumes finite value. The $j=0$ term in formulae (\ref%
{Gxxinteg1}), (\ref{Gxxinteg2}) is the Green function for the geometry with $%
b=1$ when the cosmic string is absent. Formula (\ref{Gxxinteg1}) presents
the Green function for the geometry with the cosmic string as an image sum
of the $b=1$ Green functions.

The analysis of the vacuum expectation values (VEVs) associated with a
massless scalar field in a spacetime defined by (\ref{ds2comp}) in the
absence of cosmic string has been presented in \cite{Mello0}. Here we are
mainly interested in quantum effects induced by the presence of the string.
To investigate these effects we introduce the subtracted Green function%
\begin{equation}
G_{\mathrm{sub}}(x,x^{\prime })=G(x,x^{\prime })-G(x,x^{\prime })|_{b=1}.
\label{subG}
\end{equation}%
As the presence of the string does not change the curvature for the
background manifold for the points $\rho \neq 0$, the structure of the
divergences in the coincidence limit is the same for both terms on the right
hand side. Hence, for these points the function $G_{\mathrm{sub}%
}(x,x^{\prime })$ is finite in the coincidence limit.

By using formula (\ref{Gxxinteg1}), the VEV\ of the field square is
presented as the sum%
\begin{equation}
\langle 0|\varphi ^{2}|0\rangle =\langle \varphi ^{2}\rangle _{\mathrm{m}%
}+\langle \varphi ^{2}\rangle _{\mathrm{s}},  \label{phi2int1}
\end{equation}%
where the first term on the right is the corresponding VEV in the case when
the string is absent ($b=1$) and the second term is a new contribution
induced by the cosmic string. The latter is directly obtained from the
subtracted Green function in the coincidence limit:%
\begin{eqnarray}
\langle \varphi ^{2}\rangle _{\mathrm{s}} &=&-\frac{(-i)^{d}(2\pi )^{-d/2-2}%
}{\alpha ^{m}S_{m}r^{D-1}}\sum_{j=1}^{b-1}\sum_{l=0}^{\infty }D_{l}\frac{%
Q_{\nu _{l}-1/2}^{d/2+1}(\cosh w_{j})}{(\sinh w_{j})^{d/2+1}}  \notag \\
&=&\frac{(2\pi )^{-\frac{p}{2}}\alpha ^{-m}}{2S_{m}r^{D-1}}%
\sum_{l=0}^{\infty }\frac{D_{l}\Gamma (\mu _{l})}{2^{\nu _{l}}\Gamma (\nu
_{l}+1)}\sum_{j=1}^{b-1}(\cosh w_{j})^{-\mu _{l}}F\left( \frac{\mu _{l}+1}{2}%
,\frac{\mu _{l}}{2};\nu _{l}+1;\frac{1}{\cosh ^{2}w_{j}}\right) ,
\label{phi2ints}
\end{eqnarray}%
where we have used the relation $C_{l}^{p}(1)=\Gamma (l+2p)/\Gamma (2p)l!$.
In formula (\ref{phi2ints}) the factor
\begin{equation}
D_{l}=(2l+m-1)\frac{\Gamma (l+m-1)}{\Gamma (m)\,l!}  \label{Dlang}
\end{equation}%
is the degeneracy of each angular mode with given $l$ and
\begin{equation}
\cosh w_{j}=1+2\frac{\rho ^{2}}{r^{2}}\sin ^{2}(\pi j/b).  \label{vjdef}
\end{equation}

\begin{figure}[tbph]
\begin{center}
\begin{tabular}{cc}
\epsfig{figure=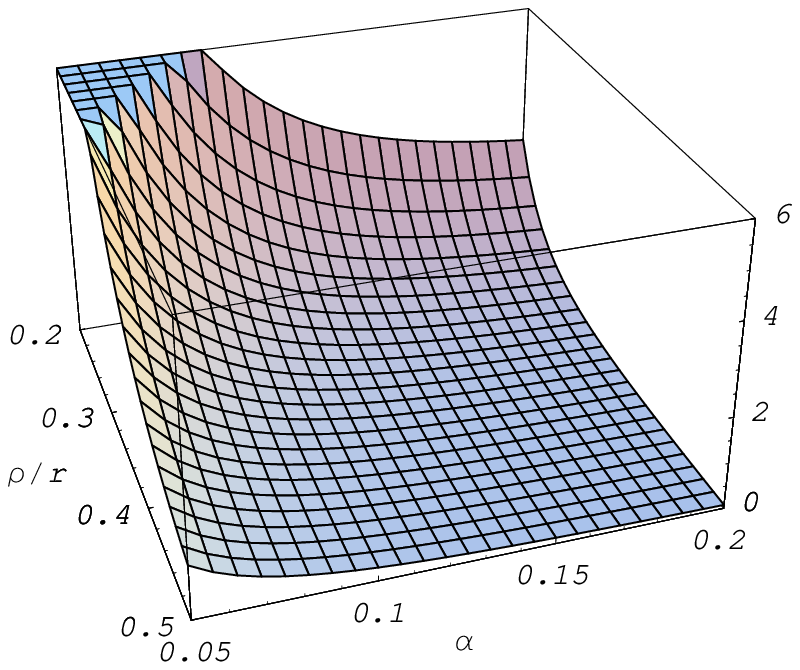, width=7cm, height=6cm} & \quad %
\epsfig{figure=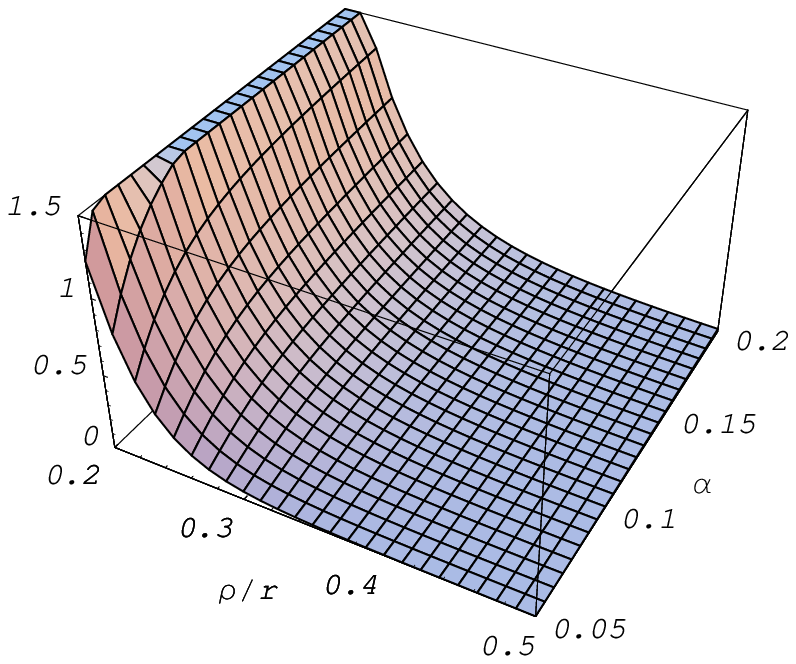, width=7cm, height=6cm}%
\end{tabular}%
\end{center}
\caption{Part in the VEV of the field square induced by the cosmic string, $%
r^{D-1}\langle \protect\varphi ^{2}\rangle _{\mathrm{s}}$, as a function on $%
\protect\rho /r$ and $\protect\alpha $ in the model with the parameters $d=1$%
, $m=2$, $b=3$. The left and right panels correspond to minimally and
conformally coupled scalar fields respectively.}
\label{fig1}
\end{figure}

In figure \ref{fig1} we have presented the VEV of the field square induced
by the cosmic string for minimally (left panel) and conformally (right
panel) coupled scalar fields as a function on $\rho /r$ and $\alpha $ in the
model with $d=1$, $m=2$, $b=3$. From this figure we see that the behavior of
the field square for small values of the parameter $\alpha $ depends
essentially on the value of the curvature coupling parameter. For a
minimally coupled scalar the VEV induced by the string increases with
decreasing $\alpha $, whereas for a conformally coupled scalar this VEV
vanishes in the limit $\alpha \rightarrow 0$. In the next section this
behavior will be analytically derived from the corresponding formulae for
the general case of the parameter $b$.

\section{General Case}

\label{sec:gencase}

\subsection{Green function}

\label{subsec:GF}

In order to evaluate the vacuum polarization effect induced by the cosmic
string for general values of the parameter $b$, we consider the function
which is obtained from the heat kernel (\ref{Kxx2}) subtracting the
corresponding heat kernel for the geometry without a cosmic string. The
latter is obtained from (\ref{Kxx2}) substituting $b=1$. By using the
notation $K_{\mathrm{sub}}(x,x^{\prime };s)$ for the subtracted function,
one finds
\begin{eqnarray}
K_{\mathrm{sub}}(x,x^{\prime };s) &=&\frac{\left( rr^{\prime }\right)
^{(1-m)/2}}{(4\pi )^{p/2}\alpha ^{m}}\frac{e^{-V/4s}}{s^{p/2+1}}%
\sum_{l=0}^{\infty }\frac{2l+m-1}{(m-1)S_{m}}C_{l}^{(m-1)/2}(\cos \theta
)I_{\nu _{l}}\left( \frac{rr^{\prime }}{2s}\right)   \notag \\
&&\times \sideset{}{'}{\sum}_{n=0}^{\infty }\left[ bI_{nb}\left( \frac{\rho
\rho ^{\prime }}{2s}\right) \cos (nb\Delta \phi )-I_{n}\left( \frac{\rho
\rho ^{\prime }}{2s}\right) \cos (n\Delta \phi )\right] .  \label{Kxxsub}
\end{eqnarray}%
To provide a more convenient expression for the subtracted heat kernel, we
apply the Abel-Plana formula (see, for instance, \cite{Most,SahaRev}) for
the summation over $n$:
\begin{equation*}
\sideset{}{'}{\sum}_{n=0}^{\infty }F(n)=\int_{0}^{\infty }du\
F(u)+i\int_{0}^{\infty }du\ \frac{F(iu)-F(-iu)}{e^{2\pi u}-1}\ .
\end{equation*}%
In this formula we take $F(u)=I_{uw}(z)\cos (uw\Delta \phi )$ with $z=\rho
\rho ^{\prime }/2s$ and $w=b,1$ for the first and second terms in the square
brackets in (\ref{Kxxsub}), respectively. Now we can see that in the
evaluation of the difference the terms coming from the first integral on the
right of Abel-Plana formula cancel out and one obtains
\begin{equation}
\sideset{}{'}{\sum}_{n=0}^{\infty }\left[ bI_{nb}(z)\cos (nb\Delta \phi
)-I_{n}(z)\cos (n\Delta \phi )\right] =\frac{2}{\pi }\int_{0}^{\infty
}du\cosh (u\Delta \phi )g(b,u)K_{iu}(z),  \label{ndif}
\end{equation}%
where $K_{\nu }(z)$ is the MacDonald function and we have introduced the
notation%
\begin{equation}
g(b,u)=\sinh (\pi u)\left( \frac{1}{e^{2\pi u/b}-1}-\frac{1}{e^{2\pi u}-1}%
\right) .  \label{gbu}
\end{equation}

The respective subtracted Green function becomes:%
\begin{eqnarray}
G_{\mathrm{sub}}(x,x^{\prime }) &=&\frac{(rr^{\prime })^{(1-D)/2}}{%
2^{p/2-1}\pi ^{p/2+1}\alpha ^{m}}\sum_{l=0}^{\infty }\frac{2l+m-1}{(m-1)S_{m}%
}C_{l}^{(m-1)/2}(\cos \theta )  \notag \\
&&\times \int_{0}^{\infty }du\,g(b,u)\cosh (u\Delta \phi )\int_{0}^{\infty
}dv\,v^{p/2-1}e^{-Vv/2rr^{\prime }}I_{\nu _{l}}\left( v\right) K_{iu}\left(
\frac{\rho \rho ^{\prime }}{rr^{\prime }}v\right) .  \label{Gxxsub2}
\end{eqnarray}%
For the points outside the cores of the topological defects, the expression
on the right of this formula is finite in the coincidence limit and can be
directly used for the evaluation of the vacuum expectation values of the
field square and the energy-momentum tensor.

\subsection{VEVs for the field square and energy-momentum tensor}

\label{subsec:emt}

In formula (\ref{Gxxsub2}), taking the coincidence limit for the part of the
VEV of the field square induced by the cosmic string one finds%
\begin{eqnarray}
\langle \varphi ^{2}\rangle _{\mathrm{s}} &=&\frac{2^{1-p/2}r^{1-D}}{\pi
^{p/2+1}\alpha ^{m}S_{m}}\sum_{l=0}^{\infty }D_{l}\int_{0}^{\infty
}du\,g(b,u)  \notag \\
&&\times \int_{0}^{\infty }dv\,v^{p/2-1}e^{-\left( 1+y\right) v}I_{\nu
_{l}}\left( v\right) K_{iu}\left( yv\right) ,  \label{phi2s1}
\end{eqnarray}%
where $y=\rho ^{2}/r^{2}$. For the case $\alpha =1$ corresponding to the
absence of the global monopole one has $\nu _{l}=l+(m-1)/2$ and the
summation over $l$ can be done explicitly by using the formula%
\begin{equation}
\sum_{l=0}^{\infty }D_{l}I_{l+\nu }(v)=\frac{e^{v}}{\nu \Gamma (\nu )}\left(
\frac{v}{2}\right) ^{\nu }.  \label{Suml}
\end{equation}%
(This formula is obtained from a more general addition theorem given in \cite%
{Pru}.) After the evaluation of the $v$-integral with the help of formula
from \cite{Pru}, one finds%
\begin{equation}
\langle \varphi ^{2}\rangle _{\mathrm{s}}|_{\alpha =1}=\frac{(2\rho )^{1-D}}{%
\pi ^{\frac{D}{2}+1}\Gamma \left( \frac{D}{2}\right) }\int_{0}^{\infty
}du\,g(b,u)\left\vert \Gamma \left( \frac{D-1}{2}+iu\right) \right\vert ^{2}.
\label{phi2alf1}
\end{equation}%
For odd values $D$ the modulus of the gamma function in this formula is
expressed via the elementary functions and the integral is explicitly
evaluated. In particular, for $D=3$ one has $|\Gamma (1+iu)|^{2}=\pi u/\sinh
(\pi u)$ and from (\ref{phi2alf1}) we obtain the well-known result \cite%
{Line87,Smit90}. The result for the case $D=5$ with the recurrence relations
for the evaluation of the higher odd values are given in \cite{Beze06}.

Now we consider the behavior of the string induced VEV $\langle \varphi
^{2}\rangle _{\mathrm{s}}$ in the asymptotic regions for the parameter $y$.
For large values of $y$, introducing in (\ref{phi2s1}) a new integration
variable $z=vy$ and expanding the integrand over $1/y$, we see that the
dominant contribution \ comes from the $l=0$ term. To the leading order one
has $\langle \varphi ^{2}\rangle _{\mathrm{s}}\sim r^{2\nu _{0}+1-m}/\rho
^{2\nu _{0}+p}$. In particular, the VEV induced by the string diverges on
the core of the global monopole for $2\nu _{0}<m-1$, is finite for $2\nu
_{0}=m-1$ corresponding to $\alpha =1$, and vanishes for $2\nu _{0}>m-1$.
For a fixed value $r$ and at large distances from the string core $\langle
\varphi ^{2}\rangle _{\mathrm{s}}$ vanishes as $1/\rho ^{2\nu _{0}+p}$. As
the VEV of the field square given by (\ref{phi2s1}) diverges on the string
core corresponding to $y=0$, in the limit $y\ll 1$ the main contribution
into the sum over $l$ comes from large values $l$ and we use the uniform
asymptotic expansion for the function $I_{\nu _{l}}\left( v\right) $. As the
next step we replace the summation over $l$ by the integration, $%
\sum_{l}D_{l}f(\nu _{l})\rightarrow (2\alpha ^{m}/\Gamma
(m))\int_{0}^{\infty }dx\,x^{m-1}f(x)$, and change the order of the
integrations. Further introducing in the integral over $x$ a new integration
variable $t=x/v$, the integral over $t$ is estimated by the Laplace method.
After the evaluation of the remained integral over $v$ using formula from
\cite{Pru}, it can be seen that to the leading order $\langle \varphi
^{2}\rangle _{\mathrm{s}}$ coincides with the corresponding result for $%
\alpha =1$ case given by (\ref{phi2alf1}). This means that near the cosmic
string the most relevant contribution to the VEV comes from the string
itself.

For small values of the parameter $\alpha $, corresponding to strong
gravitational field, and for a non-minimally coupled scalar field one has $%
\nu _{l}\gg 1$ for all values $l$. Replacing the function $I_{\nu
_{l}}\left( v\right) $ by the corresponding uniform asymptotic expansion and
noting that the dominant contribution into the integral over $v$ comes from
the values $v\sim \nu _{l}$, we can estimate this integral by the Laplace
method. In this way it can be seen that the main contribution comes from $l=0
$ term and this contribution is exponentially suppressed. In the same limit,
$\alpha \ll 1$, and for a minimally coupled scalar field one has $\nu
_{l}\gg 1$ for $l\neq 0$ terms and their contribution is exponentially
suppressed. For $l=0$ term one has $\nu _{0}=(m-1)/2$ and the corresponding
contribution to the string induced VEV of the field square behaves as $%
\langle \varphi ^{2}\rangle _{\mathrm{s}}\sim \alpha ^{m}$. As we see, in
this limit the behavior of the VEV as a function on $\alpha $ is essentially
different for minimally and non-minimally coupled scalar fields. This is
also seen from figure \ref{fig1}. The similar feature takes place for the
VEVs induced in the global monopole bulk by the presence of boundaries \cite%
{Saha03a}-\cite{Beze06b}. The suppression for the case of a non-minimally
coupled scalar field can be understood if we note that for small values $%
\alpha $ one has $R\approx 1/\alpha ^{2}r^{2}$ and the term $\zeta R$ in the
field equation acts like the mass squared term.

Now we turn to the evaluation of the VEV for the energy-momentum tensor. As
in the case of the field square, this VEV is presented in the form of the
sum of the purely global monopole and string induced parts:%
\begin{equation}
\langle 0|T_{ik}|0\rangle =\langle T_{ik}\rangle _{\mathrm{m}}+\langle
T_{ik}\rangle _{\mathrm{s}}.  \label{Tikdecomp}
\end{equation}%
The string induced part is obtained by using the formula
\begin{equation}
\langle T_{ik}\rangle _{\mathrm{s}}=\lim_{x^{\prime }\rightarrow x}\partial
_{i}\partial _{k}^{\prime }{G}_{\mathrm{sub}}(x,x^{\prime })+\left[ \left(
\xi -\frac{1}{4}\right) g_{ik}\nabla _{l}\nabla ^{l}-\xi \nabla _{i}\nabla
_{k}-\xi R_{ik}\right] \langle \varphi ^{2}\rangle _{\mathrm{s}},
\label{mvevEMT}
\end{equation}%
where $R_{ik}$ is the Ricci tensor for the background spacetime, $i,k=0$
components correspond to the standard time coordinate $t$, and the values of
the indices $i=1,2,\ldots ,D$ correspond to the coordinates $(\rho ,\phi
,z_{1},\ldots ,z_{d},r,\theta _{1},\ldots ,\theta _{m-1},\Phi )$
respectively. Note that in the first term on the right of this formula,
before the differentiation, the rotation on the time coordinate should be
made in the formula for $G_{\mathrm{sub}}(x,x^{\prime })$. From the symmetry
of the problem it follows that the vacuum stresses $\langle T_{i}^{i}\rangle
_{\mathrm{s}}$, $i=p+1,\ldots ,D$, are isotropic and one has the relations
(no summation over $i$)%
\begin{equation}
\langle T_{0}^{0}\rangle _{\mathrm{s}}=\langle T_{i}^{i}\rangle _{\mathrm{s}%
},\;i=3,\ldots ,p,  \label{T00iirel1}
\end{equation}%
between the corresponding energy density and vacuum stresses along the
string axis. From the continuity equation $\nabla _{k}\langle
T_{i}^{k}\rangle _{\mathrm{s}}=0$ we obtain the following relations for the
components of the energy-momentum tensor (no summation over $i$):%
\begin{equation}
\langle T_{2}^{2}\rangle _{\mathrm{s}}=\frac{\partial }{\partial \rho }%
\left( \rho \langle T_{1}^{1}\rangle _{\mathrm{s}}\right) ,\;\langle
T_{i}^{i}\rangle _{\mathrm{s}}=\left( 1+\frac{r}{m}\frac{\partial }{\partial
r}\right) \langle T_{p}^{p}\rangle _{\mathrm{s}},  \label{T22rel2}
\end{equation}%
with $i=p+1,\ldots ,D$. It is useful also to take into account the trace
relation%
\begin{equation}
\langle T_{i}^{i}\rangle _{\mathrm{s}}=D(\xi -\xi _{D})\nabla _{l}\nabla
^{l}\langle \varphi ^{2}\rangle _{\mathrm{s}}.  \label{trace}
\end{equation}%
For the expression appearing on the right hand side of formula (\ref{trace})
we have
\begin{eqnarray}
\nabla _{l}\nabla ^{l}\langle \varphi ^{2}\rangle _{\mathrm{s}} &=&-\frac{%
r^{-D-1}\alpha ^{-m}}{2^{p/2-2}\pi ^{p/2+1}S_{m}}\sum_{l=0}^{\infty
}D_{l}\int_{0}^{\infty }du\,g(b,u)\int_{0}^{\infty }dv\,v^{p/2-1}e^{-(1+y)v}
\notag \\
&&\times \left\{ 2\left[ 2v^{2}(1+y)-2v+\nu _{l}^{2}-u^{2}/y\right] I_{\nu
_{l}}\left( v\right) K_{iu}\left( yv\right) \right.   \notag \\
&&\left. -v\left( 4v+m-1\right) I_{\nu _{l}}^{\prime }\left( v\right)
K_{iu}\left( yv\right) -v\left( 4yv-m+1\right) I_{\nu _{l}}\left( v\right)
K_{iu}^{\prime }\left( yv\right) \right\} .  \label{Dalambphi2}
\end{eqnarray}%
By using (\ref{mvevEMT}), for the remained components of the energy-momentum
tensor we find the formulae (no summation over $i$)%
\begin{eqnarray}
\langle T_{i}^{i}\rangle _{\mathrm{s}} &=&-\frac{r^{-1-D}\alpha ^{-m}}{%
2^{p/2-1}\pi ^{p/2+1}S_{m}}\sum_{l=0}^{\infty }D_{l}\int_{0}^{\infty
}du\,g(b,u)\int_{0}^{\infty }dv\,v^{p/2-1}  \notag \\
&&\times e^{-\left( 1+y\right) v}F^{(i)}(y,u,v)+\left( \xi -\frac{1}{4}%
\right) \nabla _{l}\nabla ^{l}\langle \varphi ^{2}\rangle _{\mathrm{s}},
\label{Tiis}
\end{eqnarray}%
where $y$ is defined after formula (\ref{phi2s1}) and we have
introduced
notations%
\begin{eqnarray}
F^{(0)}(y,u,v) &=&vI_{\nu _{l}}\left( v\right) K_{iu}\left( yv\right) ,
\label{F0} \\
F^{(1)}(y,u,v) &=&\frac{1}{y^{2}}I_{\nu _{l}}\left( v\right) \left\{ -yv%
\left[ 2yv(1-4\xi )-2\xi \right] K_{iu}^{\prime }\left( yv\right) \right.
\notag \\
&&+\left. \left[ (2y^{2}v^{2}-u^{2})(1-4\xi )+2\xi yv\right] K_{iu}\left(
yv\right) \right\} ,  \label{F1} \\
F^{(p)}(y,u,v) &=&K_{iu}\left( yv\right) \left\{ 2v\left[ (\beta +v)(4\xi
-1)+\xi \right] I_{\nu _{l}}^{\prime }\left( v\right) \right.   \notag \\
&&+\left. \left[ (2v^{2}+\nu _{l}^{2}+2\beta v+\beta ^{2})(1-4\xi )+2\xi
(v-\beta )\right] I_{\nu _{l}}\left( v\right) \right\} ,  \label{FN3}
\end{eqnarray}%
with $\beta =(m-1)/2$. The consideration of the asymptotic cases of the
general formulae for the components of the energy-momentum tensor is similar
to that for the field square. For large values of $y$ the main contribution
comes from the $l=0$ term and these components behave as $r^{-D-1}(\rho
/r)^{p+2\nu _{0}}$. In the limit $y\ll 1$ corresponding to the points near
the core of the cosmic string, to the leading order the energy-momentum
tensor coincides with the corresponding quantity for the $\alpha =1$ case.
In particular, for the energy density one has%
\begin{eqnarray}
\langle T_{0}^{0}\rangle _{\mathrm{s}} &\approx &\langle T_{0}^{0}\rangle _{%
\mathrm{s}}|_{\alpha =1}=-\frac{2^{1-D}\rho ^{-1-D}}{\pi ^{\frac{D}{2}%
+1}\Gamma \left( \frac{D}{2}\right) }\int_{0}^{\infty }du\,g(b,u)  \notag \\
&&\times \left\vert \Gamma \left( \frac{D-1}{2}+iu\right) \right\vert ^{2}%
\left[ (D-1)^{2}(\xi -\xi _{D})+\frac{u^{2}}{D}\right] .  \label{T00alf1}
\end{eqnarray}%
As a special case, for $D=3$ we obtain the result given in
\cite{Frol87}. For small values of the parameter $\alpha $ the
dominant contribution to the vacuum energy-momentum tensor induced
by the cosmic string comes from the term with $l=0$ and this
tensor is exponentially suppressed for the non-minimally coupled
scalar and behaves like $1/\alpha ^{m}$ for a minimally coupled
scalar.

\section{Conclusion}

In this paper we have investigated quantum vacuum effects for a massless
scalar field induced by a composite topological defect. Specifically we have
considered the spacetime being a direct product of the cosmic string and
global monopole geometries. The corresponding heat kernel is constructed and
on the base of this the Green function is evaluated. The complete Green
function is composed by two terms: (i) the first one being singular at the
coincidence limit; this term contains information only on the global
monopole defect, and (ii) a regular term which contains information about
the presence of both topological defects and vanishes in the absence of the
cosmic string ($b=1$). First we consider the case when the parameter $b$
describing the planar angle deficit in the cosmic string geometry is an
integer. In this case the Green function is presented as an image sum of the
Green functions corresponding to the geometry which has a structure of a $p$%
-dimensional Minkowskian brane with a global monopole in transverse
dimensions. The vacuum polarization in the latter geometry is considered in
\cite{Mello0} and our main interest here are the effects induced by the
cosmic string. For the points away from the cores of the topological defects
the corresponding part in the Green function is finite in the coincidence
limit and can be directly used for the evaluation of the vacuum expectation
values of the \ field square and the energy-momentum tensor. In the special
case with integer values of the parameter $b$ the string induced part in the
VEV of the field square is given by formula (\ref{phi2ints}). The general
case of the planar angle deficit is considered in Section \ref{sec:gencase}.
By using the Abel-Plana formula for the summation over the azimuthal quantum
number, we have explicitly subtracted from the Green function the part
corresponding to the geometry when the cosmic string is absent. The
corresponding VEV of the field square is obtained in the coincidence limit
and is given by formula (\ref{phi2s1}). We have investigated the general
formula in various limiting cases for the parameters of the model. In
particular, we have shown that for small values of the parameter $\alpha $
corresponding to strong gravitational fields, the behavior of the field
square is essentially different for minimally and non-minimally coupled
fields. For points near the string core, in the leading order, the vacuum
densities coincide with those for the geometry when the global monopole is
absent corresponding to $\alpha =1$. We have also evaluated the the VEV of
the energy-momentum tensor induced by the string. The corresponding
independent components are given by formulae (\ref{Tiis})-(\ref{FN3}). Other
components can be obtained by using the continuity equation from relations (%
\ref{T22rel2}).

\section*{Acknowledgement}

AAS was supported by PVE/CAPES Program and in part by the Armenian Ministry
of Education and Science Grant No. 0124. ERBM thanks Conselho Nacional de
Desenvolvimento Cient\'\i fico e Tecnol\'ogico (CNPq) and FAPESQ-PB/CNPq
(PRONEX) for partial financial support.

\end{document}